\documentclass[
    ,final            
  ,draft            
  ,numberedheadings 
  ]
  {aipproc}

\usepackage[mathscr]{eucal}
 \usepackage{amsmath, amssymb}

\def\dmf{\dot{\mathfrak{M}}}
 
  \def\sun{\hbox{$\odot$}}
   
\newcommand{\be}{\begin{equation}}
\newcommand{\ee}{\end{equation}}
\newcommand{\bdm}{\begin{displaymath}}
\newcommand{\edm}{\end{displaymath}}
\layoutstyle{6x9}

\begin{document}

\noindent Published in {\it Astronomy Reports} {\bf 59}, 25 (2015)

\vspace{1cm}

\title{A scenario of the formation of isolated X-ray pulsars with anomalously long period}


\author{N.R.\,Ikhsanov\footnote{Saint-Petersburg State University,
Universitetsky pr., 28, Saint Petersburg 198504, Russia}}{
  address={Pulkovo Observatory, Pulkovskoe Shosse 65, Saint-Petersburg 196140, Russia}
}

\author{V.Yu.\,Kim}{
  address={Pulkovo Observatory, Pulkovskoe Shosse 65, Saint-Petersburg 196140, Russia}
}

\author{N.G.\,Beskrovnaya}{
  address={Pulkovo Observatory, Pulkovskoe Shosse 65, Saint-Petersburg 196140, Russia}
}

\begin{abstract}
A scenario of the formation of isolated X-ray pulsars is discussed with an
application to one of the best studied objects of this class 1E~161348-5055.
This moderately luminous, $10^{33} - 10^{35}\,{\rm erg\,s^{-1}}$, pulsar with a
relatively soft spectrum, $kT \sim 0.6-0.8$\,keV, is associated with an
isolated neutron star, which is located near the center of the young ($\sim
2000$\,yr) compact supernova remnant RCW\,103 and rotates steadily
($|\dot{\nu}| \leq 2.6\,\times\,10^{-18}\,{\rm Hz\,s^{-1}}$) with the period of
6.7\,hr. We show that in the current epoch the neutron star is in the accretor
state. The parameters of the source emission can be explained in terms of the
magnetic-levitation accretion scenario in which the star with the surface
magnetic field of $10^{12}$\,G accretes material onto its surface from a
non-Keplerian magnetic fossil disk at the rate $10^{14}\,{\rm g\,s^{-1}}$. A
neutron star could evolve to this state in a High-Mass X-ray Binary (HMXB),
which had disintegrated during the supernova explosion powered by the
core-collapse of its massive component. The life-time of an isolated X-ray
pulsar formed this way can be as long as a few thousand years.
\end{abstract}

\maketitle


   \section{Introduction}

Isolated X-ray pulsars  constitute a subclass of isolated neutron stars (i.e.
not associated with a close binary system)  displaying regular pulsations in
their X-ray emission. This subclass now includes more than 70 objects. Among
them are radio-pulsars emitting also X-rays (see \cite{Malov-2004} and
references therein), anomalous X-ray pulsars and soft gamma-ray repeaters  (see
\cite{Bisnovatyi-Kogan-Ikhsanov-2014} and references therein), compact X-ray
sources in supernova remnants \cite{Pavlov-etal-2004}, and isolated neutron
stars of a relatively low luminosity, known under a romantic name ``The
Magnificent Seven'' \cite{Treves-etal-2001}. A great diversity in observational
manifestations of these stars reflects differences in mechanisms responsible
for generation of their X-ray emission. Nevertheless, assignment of these
objects to a single sub-class is justified by a similarity of their rotational
evolution. The average spin period of these stars,  $P_{\rm s}$, is
monotonously increasing,   $\dot{P} >0$, and their spin-down ages, $\tau \sim
P_{\rm s}/2\,\dot{P}$, are mostly in the range  $10^3 - 10^5$\,yr (see
\cite{Malov-Machabeli-2009} and references therein).

The most simple explanation of the observed spin evolution of isolated X-ray
pulsars is presented by the canonical model of a radio-pulsar in which the
spin-down power of the star is described by  the magneto-dipole formula,
$L_{\rm md} = f_{\rm m} \mu^2 \omega_{\rm s}^4/c^3$, where $\omega_{\rm s} =
1/P_{\rm s}$ is the angular velocity and $\mu = (1/2) B_* R_{\rm ns}^3$ is the
dipole magnetic moment of the neutron star of radius  $R_{\rm ns}$, with the
surface magnetic filed  $B_*$. The dimensionless parameter  $f_{\rm m}$ ranges
within $1 \leq f_{\rm m} \leq 4$ according to \cite{Spitkovsky-2006,
Beskin-2010}. In this approach, the isolated X-ray pulsars are described in
terms of a young neutron star with strong magnetic field. A number of
mechanisms of X-ray emission from these objects have been considered. Among
them are current dissipation and particle acceleration in the pulsar
magnetosphere, dissipation of the super-strong magnetic field in the neutron
star crust, and heating of the stellar photosphere by the thermal energy coming
from its hot core (cooling), which can proceed at a  rather high rate during
the early stage of stellar evolution  \cite{Hansel-etal-2007}.

It should be noted, however, that the radio-pulsar model is not an universal
instrument for analysis of spin evolution of neutron stars. In particular, it
can be applied only to those pulsars whose period satisfies the condition
$P_{\rm s} < P_{\rm ej}$, where
  \be\label{pej}
 P_{\rm ej} \simeq 20\ f_{\rm m}^{1/4}\,\mu_{30}^{1/2}\,m^{-1/2}\,n^{-1/4}\,v_7^{1/2}\ \text{s}
 \ee
is a period at which the star switches its state from ejector (i.e., a
spin-powered pulsar) to propeller  \cite{Popov-etal-2000}. Here $\mu_{30} =
\mu/10^{30}\,\text{G\,$\cdot\,$cm$^3$}$ and $m$ is the neutron star mass,
$M_{\rm ns}$, in units of $1.4\,{\rm M_{\sun}}$. $n$ is the gas density around
the pulsar magnetosphere, and  $v_7$ is the relative velocity between the star
and the surrounding material,  $v_{\rm rel}$, in units of  $10^7$\,cm/s.
Assuming, that the gas density around
 the star at the current stage of its evolution does not significantly
deviate from the number density  in the interstellar space (i.e. about one atom
per cubic centimeter), we can see that the above-mentioned condition is
satisfied for the majority of currently known isolated X-ray pulsars. The only
exception is the X-ray pulsar   1E~161348-5055 (hereafter 1E~1613) with the
period $\sim 6.7$\,hr. Analyzing the parameters of this pulsar in
Section\,\ref{parameters}, we come to the conclusion that it is a neutron star
in the accretor state (Section\,\ref{evol-state}). Good agreement with
observational data can be achieved in the scenario of magnetic-levitation
accretion, in which a neutron star with magnetic field  $\sim 10^{12}$\,G
accretes material onto its surface from a non-Keplerian magnetized residual
disk. Magneto-rotational evolution of this star and the origin of its fossil
disk are discussed in Section\,\ref{origin} under assumption that 1E~1613 is a
descendant of a binary system. We show that the origin of this source can be
explained within a canonical evolutionary model for a High-Mass X-ray Binary
(HMXB), which had disintegrated after the supernova explosion caused by a
core-collapse of its massive component. The pulsar
 1E~1613 in this scenario is an old neutron star which had been born in the first supernova
 explosion and evolved as a part of a HMXB. Summarizing main conclusions of our research in
 Section \,\ref{discussions}, we estimate parameters of the binary systems which could be
 progenitors of the long-period isolated X-ray pulsars.

 \section{Parameters of  1E~161348-5055}\label{parameters}

The point-like X-ray source  1E~1613 was discovered in  1979 on board of the
Einstein space mission  through observations of the supernova remnant  RCW\,103
\cite{Tuohy-Garmire-1980}. The source is located close to the center of the
nebulosity, the distance to which is  $ 3.2 \pm 0.1$\,kpc
\cite{Caswell-etal-1975}. This nebula is not a typical remnant of SN\,II
explosion. It is round-shaped, has a fibrous structure, low expansion velocity
($\sim \, 1100$\,km/c) and, for its age,   $\tau_0 \simeq 2000 \pm 1000$\,yr,
relatively small spatial dimensions  $\sim 7.7$\,pc \cite{Carter-etal-1997}.
Nebulosities with such parameters constitute less than  20\% of all known by
now remnants of  SN\,II, which probably exploded in a gaseous medium of
enhanced density \cite{Marsden-2001}.

The first doubts concerning the assumption that the source  1E~1613 is a young
cooling neutron star arose after spectral observations of the object with the
space observatory ASCA \cite{Gotthelf-etal-1997}. Its  X-ray emission was found
to have a mean luminosity   $L_{\rm X} \approx 10^{34}$\,erg/s, and to contain
a blackbody component of temperature  $kT \sim 0.6-0.8$\,keV and emitting
radius   $a_{p} \sim 600$\,m, which is substantially smaller than the radius of
the star itself  \cite{Gotthelf-etal-1997, Gotthelf-etal-1999}. Moreover,
analyzing the data on 1E~1613 from three space telescopes (Einstein, ASCA and
ROSAT), Gotthelf et al. \cite{Gotthelf-etal-1999} have argued that the X-ray
brightness of the source undergo slow variations with the amplitude reaching
the order of magnitude, that is untypical for a cooling neutron star, but is an
attribute of neutron stars accreting material onto their surface.

Brightness variations of  1E~1613 with the period  $\sim 6$\,hr, suspected in
Chandra observations of this source, stimulated the efforts in search for an
optical counterpart to this object. Observations made with  ESO VLT in the near
infra-red allowed to set an upper limit to the luminosity of a hypothetical
companion,   $L_{\rm IR} \leq 4 \times 10^{31}$\,erg/s, which could be only a
star later than $M4$ \cite{Pavlov-etal-2004}. This result argued in favor of
assertion that  1E~1613 is not a member of any binary system, i.e. it is an
isolated neutron star emitting due to either accretion from a fossil disk or
rapid dissipation of the  super-strong magnetic field  \cite{Li-2007}.

The period of pulsations, $P_{\rm obs} = 6.67 \pm 0.03$\,hr, was measured
through X-ray observations of  1E~1613 obtained with XMM-Newton
\cite{De-Luca-etal-2006}. Discussion on the evolutionary status of the pulsar
inspired  by this discovery resulted in a number of scenarios in which  1E~1613
was considered in terms of a neutron star with super-strong magnetic field ($
\geq 10^{15}$\,G) which either accretes material from a residual Keplerian disk
\cite{De-Luca-etal-2006, Li-2007}, or is a member of a low-mass X-ray binary
\cite{Pizzolato-etal-2008}. A hypothesis about 1E~1613 being a millisecond
pulsar in a low-mass close binary system with the orbital period of $6.7$\,hr
was discussed in  \cite{Bhadkamkar-Ghosh-2009}.

High stability of the pulse period, $|\dot{P}| \leq
1.6\,\times\,10^{-9}\,\text{s/s}$ (or, correspondingly, $|\dot{\nu}| \leq 2.8
\times 10^{-18}$\,Hz/s, where $\nu = 1/P_{\rm s}$ is the spin frequency),
detected by Esposito et\,al. \cite{Esposito-etal-2011} through analysis of
X-ray observations of  1E~1613 with Swift, Chandra and XMM-Newton, has opened a
new era in the study of this exotic object. Examining their results, the
authors of this discovery pointed out that the torque applied to the neutron
star at the present time is significantly smaller than the value expected in
all previously suggested  models of this source. High stability of the
pulsations shows that its spin period is now close to the equilibrium period
whose value only slightly depends on variations of accretion rate from a
residual disk. In the previous paper  \cite{Ikhsanov-etal-2013} we argued that
such a situation could be realized in the scenario of magnetic-levitation
accretion in which a neutron star with magnetic field   $\sim 10^{12}$\,G
accretes onto its surface from a residual non-Keplerian magnetized disk.
Additional justification and further development of this approach is a subject
of our investigation. Its results are presented in the following sections of
this paper.

 \section{Evolutionary state of 1E~1613}\label{evol-state}

One of the key problems in the modeling of 1E~1613 is the current evolutionary
state of this source. Studies of spin evolution of pulsars unambiguously
indicate that the initial spin period of neutron stars at the time of birth,
$P_0$, amounts to fractions of a second \cite{Narayan-1987, Noutsos-etal-2013}.
If $P_0 < P_{\rm ej}$, the star begins its evolution in the ejector state in
which the spin period, $P_{\rm s}$, grows according to the canonical model of
radio-pulsars. During the phase when  $P_{\rm ej} < P_{\rm s} < P_{\rm pr}$,
the rotational  power of the star decreases  through the propeller mechanism.
Here $P_{\rm pr}$ determines the value at which the corotation radius of the
neutron stars, $r_{\rm cor} = \left(GM_{\rm ns}/\omega_{\rm s}^2\right)^{1/3}$,
reaches its magnetosphere radius, $r_{\rm m}$. If $P_{\rm s} \geq P_{\rm pr}$,
the star is in the accretor state, in which evolution of its spin frequency is
governed by the equation
 \be\label{spin}
 2 \pi I \dot{\nu} = K_{\rm su} - K_{\rm sd}.
 \ee
Here $I$ is the moment of inertia of the neutron star, $\dot{\nu} = d\nu/dt$,
and $K_{\rm su}$ and $K_{\rm sd}$ are the spin-up and spin-down torques exerted
on the star from the accretion flow. Hence, the neutron star can be observed in
one of three states: ejector, propeller or accretor  \cite{Shvartsman-1970,
Illarionov-Sunyaev-1975, Lipunov-1987}.

A possibility that 1E~1613 is presently in the ejector state can be excluded.
The inequality $P_{\rm obs} < P_{\rm ej}$ can be satisfied for the parameters
of this source only if the surface magnetic field of the neutron star is in
excess of  $10^{18}$\,G. However, the existence of a neutron star with so
strong magnetic field seems highly improbable  (see \cite{Potekhin-2010} and
references therein). Moreover, the expected spin-down rate of such an ejector,
 \be
\dot{\nu}_{\rm ej} = \frac{L_{\rm md}}{2 \pi I \omega_{\rm s}} = \frac{f_{\rm
m} \mu^2 \omega_{\rm s}^3}{2 \pi I c^3},
 \ee
exceeds the upper limit of the period derivative of 1E~1613, derived in
\cite{Esposito-etal-2011}, by two orders of magnitude.

 An assumption that the star is in
the propeller state means that its spin period satisfies the condition  $P_{\rm
ej} < P_{\rm s} < P_{\rm pr}$, where (see Eq.\,22 from \cite{Ikhsanov-2007}),
    \be\label{pprsp}
P_{\rm pr} = P_{\rm pr}^{\rm (sp)} \simeq 10^3\
\mu_{30}^{6/7}\,m^{-11/7}\,n^{-3/7}\,v_7^{9/7}\ \text{s}.
   \ee
A characteristic spin-down time of the star in this state is  (see Eq.\,21 from
\cite{Ikhsanov-2007})
 \be\label{tauprsp}
 \tau_{\rm pr}^{\rm (sp)} \simeq 2 \times 10^{11}\ \mu_{30}^{-1}\,I_{45}\,m^{-1}\,n^{-1/2}\ \text{yr},
 \ee
where $I_{45} = I/10^{45}\,\text{g\,cm$^2$}$. Solving the system of equations
~(\ref{pprsp}) and (\ref{tauprsp}) for $\mu$, we find that the spin period of
the star in the propeller state could reach the value $P_{\rm pr}^{\rm (sp)}
\sim 6.7$\,hr on a time span of $\tau_{\rm pr}^{\rm (sp)} \sim 2000$\,yr, only
if $\mu \geq \mu_{\rm pr}$, where
 \be
 \mu_{\rm pr} \simeq 6 \times 10^{34}\ \text{G\,$\cdot$\,cm$^3$}\ \times\ I_{45}^{1/2}\,m^{5/12}\,v_7^{3/4}\
 \left(\frac{\tau_{\rm pr}^{\rm (sp)}}{2000\,\text{yr}}\right)^{-1/2}
 \left(\frac{P_{\rm pr}^{\rm (sp)}}{6.7\,\text{hr}}\right)^{7/12}.
 \ee
The magnetic field strength on the stellar surface in this case is in excess of
$10^{17}$\,G, and the star is spinning down at the rate exceeding the value
reported in  \cite{Esposito-etal-2011} by seven orders of magnitude. This
finding rules out a possibility for the neutron star in  1E~1613 to be in the
propeller state.

Above mentioned discrepancies force us to conclude that the neutron star in
1E~1613 is currently in the accretor state. In this case its X-ray emission can
be described in terms of matter infall at the rate  \be\label{dmf0} \dmf_0
\simeq 5 \times 10^{13}\ m^{-1}\ R_6\ L_{34}\ \text{g/s}
 \ee
onto the surface of a neutron star with magnetosphere radius
  \be\label{r0}
 r_{\rm mag} \simeq 3 \times 10^8\ R_6^3\ \left(\frac{a_{\rm p}}{600\,\text{m}}\right)^{-2}\ \text{cm}.
 \ee
Here $L_{34}$ is the average X-ray luminosity of the  pulsar in units of
$10^{34}$\,erg/s, $R_6$ is the neutron star radius, $R_{\rm ns}$, in units
$10^6$\,cm, and $a_{\rm p} \approx R_{\rm ns} \left(R_{\rm ns}/r_{\rm
mag}\right)^{1/2}$ is the average radius of the hot spots at the base of the
accretion column estimated in \cite{Gotthelf-etal-1999} through the X-ray
spectrum of the source.

The  lack  of observational evidence for the binary nature of the object
suggests that the only possible source of matter to be accreted by the neutron
star is a fossil disk formed after the supernova explosion. Analyzing such
situation in the previous paper,
 \cite{Ikhsanov-etal-2013}, we have shown that  steady rotation of the star
 ($|\dot{\nu}| \leq 2.8 \times 10^{-18}$\,Hz/s) with the period  of 6.7\,hr
can be explained assuming that either the angular velocity of matter in the
residual disk is significantly smaller than the Keplerian velocity, or magnetic
field on the stellar surface is in excess of  $\sim 10^{16}$\,G. The
magnetosphere radius of the star in this case turns out to exceed by a factor
of 1000 the value $r_{\rm mag}$, derived from observations.

The best agreement with observational data can be obtained in the model of
magnetic-levitation accretion from a non-Keplerian magnetic residual disk in
which the matter is confined by its own magnetic field
\cite{Bisnovatyi-Kogan-Ikhsanov-2014, Ikhsanov-etal-2014}. The outer radius of
this disk is determined by the Shvartsman radius \cite{Shvartsman-1971},
 \be\label{rsh}
 R_{\rm sh} = \beta_0^{-2/3} r_{\!_{\rm G}} \left(\frac{c_{\rm s}(r_{\!_{\rm G}})}{v_{\rm rel}}\right)^{4/3},
 \ee
at which the magnetic pressure in the initially quasi-spherical flow reaches
the value of its ram pressure.  Here $\beta_0$ is the ratio of the thermal,
$\rho c_{\rm s}^2$, to magnetic, $B_{\rm f}^2/8 \pi$, pressure in the material
captured by the neutron star at its Bondi radius,  $r_{\!_{\rm G}} = 2 GM_{\rm
ns}/v_{\rm rel}^2$, and parameters $\rho$, $B_{\rm f}$ and $c_{\rm s}$  denote
the density, magnetic field strength and the speed of sound in the accretion
flow, respectively. At the Shvartsman radius the quasi-spherical flow is
decelerated by its own magnetic field and transforms into a slowly rotating
magnetically-levitating disk (ML-disk) \cite{Bisnovatyi-Kogan-Ruzmaikin-1974,
Bisnovatyi-Kogan-Ruzmaikin-1976}.

The inner radius of the ML-disk is expressed as  \cite{Ikhsanov-etal-2014a}
  \be\label{rma}
 r_{\rm ma} \simeq 3 \times 10^8\ \times\ \alpha_{0.1}^{2/13}\ \mu_{30}^{6/13} m^{5/13} T_6^{-2/13}
 \dmf_{14}^{-4/13}\ \text{cm},
 \ee
where $\alpha_{0.1} = \alpha/0.1$ is the ratio  of the effective coefficient of
the accretion flow diffusion into the magnetic field of the star at its
magnetosphere boundary, $D_{\rm eff}$, to the Bohm diffusion coefficient
normalized following the results presented in  \cite{Gosling-etal-1991}. $T_6$
is the plasma temperature at the magnetosphere boundary in units of  $10^6$\,K,
and $\dmf_{14}$ is the mass accretion rate onto the stellar surface in  units
of $10^{14}$\,g/s. The radius  $r_{\rm ma}$ corresponds to the radius of pulsar
magnetosphere derived from observations if the magnetic field strength on the
neutron star surface is   $B_* \sim 10^{12}$\,G.

The torque applied to the star from the ML-disk \cite{Ikhsanov-etal-2013,
Ikhsanov-etal-2014a},
 \be\label{ksds0}
 K_{\rm sl} = \frac{k_{\rm t} \mu^2}{\left(r_{\rm ma} r_{\rm cor}\right)^{3/2}} \left(1 -
 \frac{\Omega_0}{\omega_{\rm s}}\right),
 \ee
strongly depends on the angular velocity of matter at its inner radius,
$\Omega_0$. Limitation of the rate of period changes obtained from observations
\cite{Esposito-etal-2011}, can be achieved in this case under condition
$\Omega_0 \sim \omega_{\rm s}$. Moreover, using the Eq.~(\ref{ksds0}) we find
that the characteristic spin-down time-scale in the previous epoch  (i.e. when
$\Omega_0 \ll \omega_{\rm s}$),
  \be\label{taua}
 \tau_{\rm a} \simeq \frac{P_{\rm s}}{2 \dot{P}_{\rm sl}} =  \frac{I (GM_{\rm ns})^{1/2} r_{\rm ma}^{3/2}}{2 \mu^2},
 \ee
for parameters of 1E~1613 is \cite{Ikhsanov-etal-2013}
  \be
 \tau_{\rm a} \simeq 1880\ \mu_{30}^{-17/13} m^{8/13} I_{45} T_6^{3/13} \dmf_{14}^{-6/13}\,\text{yr},
 \ee
and, correspondingly, does not exceed the age of the supernova remnant RCW\,103
\cite{Carter-etal-1997}. This result opens an opportunity to explain this
exotic object without the assumption about a super-strong magnetic field of the
neutron star to be invoked. In particular, we can envisage a situation, in
which during the last 2000 years, the neutron star  has been in the state of
accretion from a slowly rotating,  $\Omega_0 \sim \omega_{\rm s}$, residual
ML-disk with initial mass   $M_{\rm d}$, which was sufficient to maintain the
process of accretion at the average rate  $\dmf_0 \sim 10^{14}\,\text{g/s}$
over a time interval of  $\tau_0 \sim 2000$\,yr, i.e. $M_{\rm d} \geq M_0$,
where
 \be
 M_0 = \dmf_0 \tau_0 \simeq 3 \times 10^{-9}\,{\rm M_{\sun}}.
 \ee

Analysis of this scenario leads us to the necessity to answer the following two
questions:
 i)~when and why a massive residual disk was formed around the neutron star,  and ii)~why the neutron star
 switched to the accretor state immediately after the supernova explosion which had given birth to the nebulosity  RCW\,103.
It is very difficult to answer the second question in the frame of evolutionary
scenario for an isolated neutron star since it requires assumption that the
spin period of the neutron star at the moment of its birth was in excess of a
critical value (see Eq.~22 in \cite{Ikhsanov-etal-2013}) \be
 P_{\rm ma} \simeq 3.5\,\text{s}\ \times\ \mu_{30}^{9/13}\,m^{-5/13}\,T_6^{-3/13}\,\dmf_{14}^{-6/13},
 \ee
at which the equality $r_{\rm ma} = r_{\rm cor}$ is valid. However, we cannot
exclude a possibility that the pulsar  1E~1613 is a descendant of a HMXB and
was born long before the explosion of its massive companion resulted in
formation of supernova remnant  RCW\,103. Discussing this possibility in the
next Section, we show that by the moment of the second supernova explosion the
spin period of the old neutron star could significantly exceed the value
$P_{\rm ma}$ and the mass of disk surrounding its magnetosphere under certain
conditions could be comparable or even exceed  $M_0$. Accreting from a residual
disk, the star will manifest itself as an X-ray pulsar even after the
disruption of a binary system on a time span of at least a few thousand years.

 \section{Formation of an isolated pulsar due to binary disruption}\label{origin}
We consider  a situation in which   1E~1613 is a neutron star which was born in
a close massive binary system in the first supernova explosion. We assume that
the system was not disintegrated and a neutron star formed a close pair with
its massive companion. The initial spin period of the neutron star was of the
order of fractions of a second and then steadily grew as the star had passed
the ejector and propeller phases. Afterwards the star had switched to the
accretor state, in which it remained until the second supernova explosion,
caused by the core-collapse of its massive companion. This event was very
likely to result in system disintegration \cite{Popov-Prokhorov-2006}, and the
old neutron star became an isolated pulsar embedded in the supernova remnant of
its companion and observed now in a form of nebulosity  RCW\,103.

 \subsection{Magneto-rotational evolution}

By now, the above-described evolutionary scenario for a HMXB has been quite
adequately studied and is considered as canonical by the majority of authors
(see \cite{Postnov-Yungelson-2014} and references therein). The life-time of
such a system is limited by the evolutionary time of its optical component on
the main sequence which for the star of mass  $M_{\rm opt}$ is
\cite{Bhattacharya-van-den-Heuvel-1991}
 \be
 t_{\rm ms} \simeq 6 \times 10^6\ \left(\frac{M_{\rm opt}}{20\,{\rm M_{\sun}}}\right)^{-5/2}\ \text{yr}.
 \ee
Numerical simulations of magneto-rotational evolution of a neutron star within
this scenario with account for dissipation of its magnetic field were made by
Urpin et\,al. \cite{Urpin-etal-1998}. According to their results, the stellar
magnetic field decreases by an order of magnitude on the average due to
diffusion and accretion screening \cite{Bisnovatyi-Kogan-Komberg-1974} over a
time interval  $t_{\rm ms}$. In particular, the initial surface magnetic field
of the star $\sim 10^{13}$\,G will decrease down to $\sim 10^{12}$\,G in the
course of its evolution as a member of a HMXB. Throughout most of its
evolution, the star remains in the ejector state (see Eq.\,17 from
\cite{Ikhsanov-2012})
  \be\label{tauej}
\tau_{\rm ej} \sim 3 \times 10^6\
\mu_{30.5}^{-1}\,I_{45}\,\dmf_{14}^{-1/2}\,v_7^{-1/2}\,\text{yr}.
 \ee
During this time its magnetic field drops by a factor of 3 (rapid cooling of
the star during this phase significantly reduces the rate of  magnetic field
diffusion in its crust  \cite{Urpin-etal-1998}).

As soon as the spin period of the star reaches its critical value  (see Eq.~16
from \cite{Ikhsanov-2012}),
  \be\label{pmd}
P_{\rm ej} \sim 1.2\,\text{s}\ \times\
\mu_{30.5}^{1/2}\,\dmf_{14}^{-1/4}\,v_7^{-1/4},
 \ee
it switches to the propeller state in which the spin-down torque exerted on the
star by the gas, penetrating inside  its Bondi radius, is given by the
expression   $K_{\rm sd}^{\rm (pr)} \sim \mu^2/r_{\rm m}^3$
\cite{Davies-Pringle-1981, Lipunov-1987}. The duration of the propeller phase,
 \be\label{taupr}
 \tau_{\rm pr} = \frac{\pi I r_{\rm m}^3}{\mu^2 P_{\rm ej}},
 \ee
strongly depends on the geometry and physical parameters of the accretion
flow.This parameter takes the minimum possible value,
 \be\label{tauprsl}
 \tau_{\rm pr}^{\rm (sl)} \simeq 6000\,\text{yr}\ \times\
 \alpha_{0.1}^{6/13}\,\mu_{30}^{-29/26}\,I_{45}\,m^{3/13}\,T_6^{-6/13}\,\dmf_{14}^{-35/52}\,v_7^{1/4},
 \ee
within the scenario of magnetic-levitation accretion, in which the star is
surrounded by a non-Keplerian ML-disk and its magnetosphere radius is $r_{\rm
m} = r_{\rm ma}$. Eq.~(\ref{tauprsl}) has been obtained by solving the system
of equations~(\ref{rma}), (\ref{pmd}) and (\ref{taupr}).

 \subsection{Parameters of a magnetically-levitating disk}

Formation of a non-Keplerian magnetic disk in the Roche lobe of the neutron
star which is a member of a HMXB can proceed if  $R_{\rm sh} > \max\{r_{\rm A},
r_{\rm circ}\}$, where $r_{\rm circ} = \xi^2\,\Omega_{\rm orb}^2\,r_{\rm
G}^4/GM_{\rm ns}$ is the circularization radius, and  $\Omega_{\rm orb} = 2
\pi/P_{\rm orb}$ is the angular velocity of the orbital motion with the period
$P_{\rm orb}$. This is valid if the neutron star velocity relative to the wind
of its companion satisfies the inequality  $v_{\rm kd} < v_{\rm rel} < v_{\rm
ma}$, where \cite{Ikhsanov-etal-2013}
   \be\label{vma}
 v_{\rm ma} \simeq 380\,\text{km/s}\ \times\ \beta_0^{-1/5} m^{12/35}
 \mu_{30}^{-6/35} \dmf_{14}^{3/35} c_6^{2/5}
 \ee
and
 \be
 v_{\rm kd} \simeq 60\,\text{km/s}\ \times\ \xi_{0.2}^{3/7}\,\beta_0^{1/7}\,m^{3/7}\,c_6^{-2/7}\,\left(\frac{P_{\rm orb}}{100\,\text{days}}\right)^{-3/7}.
 \ee
Under these conditions, over a time span of $\tau_{\rm ej} + \tau_{\rm pr}$,
the spin period of a neutron star reaches a critical value (see Eq.~22 from
\cite{Ikhsanov-etal-2013}), \be
 P_{\rm pr} \simeq 3.5\,\text{s}\ \times\  \mu_{30}^{9/13}\,m^{-5/13}\,T_6^{-3/13}\,\dmf_{14}^{-6/13},
 \ee
at which the centrifugal barrier at the magnetosphere boundary fails to prevent
accretion from a non-Keplerian ML-disk onto the stellar  surface.

The mass of matter forming a ML-disk, surrounding the neutron star
magnetosphere, can be expressed as follows:
 \be\label{mdisk}
 M_{\rm d} = 4 \pi \int\limits_{r_{\rm ma}}^{R_{\rm sh}} \rho(r) h_{\rm z}(r) r dr,
 \ee
where $\rho(r)$ and $h_{\rm z}(r)$ are the density of matter in the disk and
its half-thickness. These parameters can be evaluated taking into account that
the gaseous (as well as magnetic) pressure in the disk reaches the maximum
value,
  \be
 \rho(r_{\rm ma}) c_{\rm s}^2(r_{\rm ma}) = \frac{\mu^2}{2 \pi r_{\rm ma}^6},
 \ee
at its inner radius, corresponding to the magnetosphere radius of the star,
$r_{\rm ma}$, and decreases with distance from the magnetosphere boundary as
$\rho(r) c_{\rm s}^2(r) \propto r^{-5/2}$
\cite{Bisnovatyi-Kogan-Ruzmaikin-1974, Bisnovatyi-Kogan-Ruzmaikin-1976}. Given
that the gas temperature in the disk is
 \be\label{tdisk}
 T(r) = \left(\frac{\dmf GM_{\rm ns}}{4 \pi r^3 \sigma_{\!_{\rm SB}}}\right)^{1/4},
 \ee
and, correspondingly, the sound speed is  $c_{\rm s} \sim \left(k_{\rm B}
T/m_{\rm p}\right)^{1/2} \propto r^{-3/8}$, the density of matter in the disk
can be estimated as follows
 \be
 \rho(r) = \rho(r_{\rm ma}) \left(\frac{r}{r_{\rm ma}}\right)^{-7/4},
 \ee
where $\sigma_{\!_{\rm SB}}$ is the Stefan-Boltzmann constant,   $k_{\rm B}$ is
the Boltzmann constant and $m_{\rm p}$ is the proton mass.

The half-thickness of the disk can be expressed as
\cite{Bisnovatyi-Kogan-Ruzmaikin-1974, Bisnovatyi-Kogan-Ruzmaikin-1976}
 \be\label{hz}
 h_{\rm z}(r) = \left(\frac{k_{\rm B} T(r) r^3}{m_{\rm p} GM_{\rm ns}}\right)^{1/2}.
 \ee

Substituting (\ref{rsh}) and (\ref{tdisk}--\ref{hz}) into (\ref{mdisk}) and
bearing in mind that under the conditions of interest $R_{\rm sh} \gg r_{\rm
ma}$, we find
 \be
 M_{\rm d} \simeq 2 \times 10^{-10}\,{\rm M_{\sun}}\ \times\ \alpha_{0.1}^{-7/3}
 \beta_0^{-11/12} \mu_{30}^{5/13}\ \dmf_{14}^{99/104}\ m^{25/52} c_6^{11/6} v_7^{-55/12}.
 \ee

 \begin{figure}
 \includegraphics[scale=0.6]{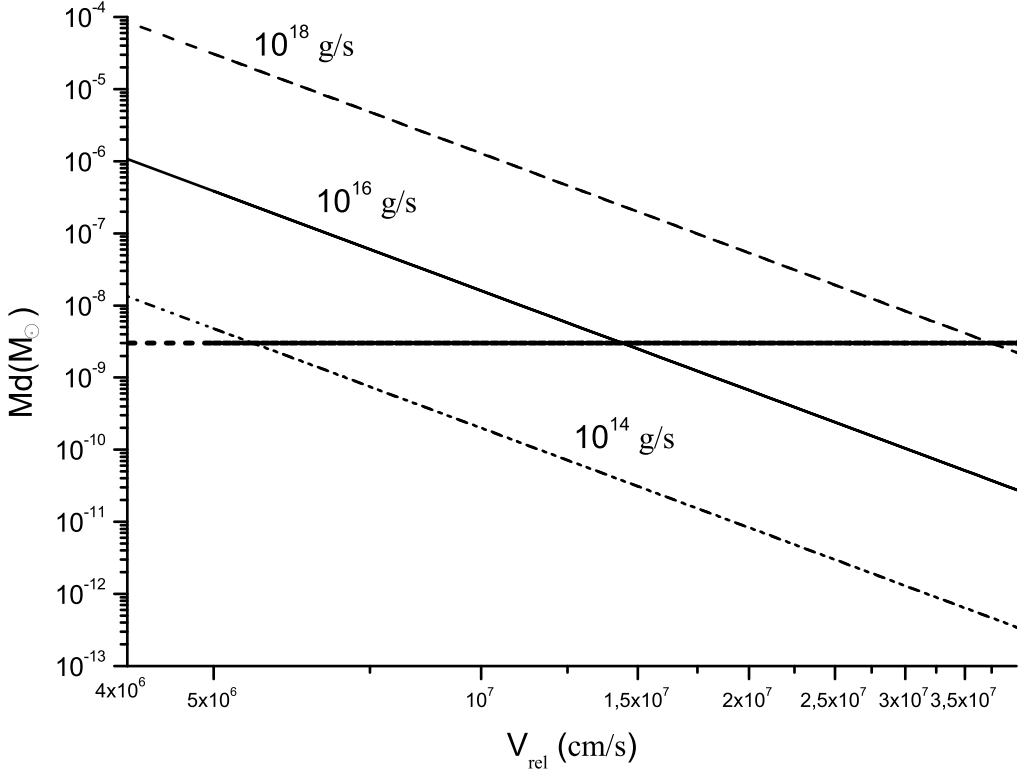}
 \caption{Dependence of the mass of a residual ML-disk, $M_{\rm d}$,
 on the relative velocity of a neutron star,  $v_{\rm rel}$, capturing material
 from the magnetized stellar wind in a high-mass X-ray binary, for
different values of the mass exchange rate, $\dmf$, and the sound speed value
in the wind $c_{\rm s}(r_{\!_{\rm G}}) = 10$\,km/s. Horizontal line indicates
the mass of a residual disk $M_0 = 3 \times 10^{-9}\,{\rm M_{\sun}}$,
necessary to maintain the process of accretion in the isolated pulsar  1E~1613
at the average rate of $10^{14}$\,g/s over a time span of 2000\,years.}
  \label{f1}
 \end{figure}

Analyzing the function $M_{\rm d} = M_{\rm d}(v_{\rm rel})$, presented in the
figure\,\ref{f1} for different values of  $\dmf$, we find that a residual disk
can be formed only in the wide systems with slow stellar wind in which $v_{\rm
rel} \sim 100$\,km/s. Such situation can be realized in the systems where a
massive component is either  Oe/Be star with outflowing disk or a red giant. In
this case the value of  $v_{\rm rel}$ is comparable to the orbital velocity of
the neutron star.  The mass of residual disk also substantially depends on the
mass exchange rate between the system components at the end of the system
evolution. In particular, for the most plausible values of   $v_{\rm rel} \sim
100 - 200$\,km/s (the orbital velocity of the neutron star in a HMXB with the
orbital period of  $100-200$\,days), the mass of a disk surrounding   1E~1613
exceeds $M_0$ if during the previous epoch, the neutron star was in the state
of accretion from a magnetized stellar wind with the rate  $\dmf \geq 10^{15} -
10^{16}$\,g/s. As has been recently shown in  \cite{Ikhsanov-etal-2014a}, the
angular velocity of matter in the magnetic disk can be evaluated using the
expression  (see also \cite{Bisnovatyi-Kogan-1991})
 \be
 \omega_{\rm sl} \sim \frac{2 \pi}{P_{\rm orb}} \left(\frac{r_{\!_{\rm G}}}{R_{\rm sh}}\right)^2,
 \ee
which assumes that inside the region  $R_{\rm sh} \leq r \leq r_{\!_{\rm G}}$,
accretion proceeds in a quasi-spherical regime with conservation of the angular
momentum, while a magnetic disk, in which the movement of material is
controlled by the intrinsic magnetic field of the  flow, is in  solid-body
rotation. The condition $\omega_{\rm sl} \sim \omega_{\rm s}$ in this case is
satisfied if  $c_{\rm s}(r_{\!_{\rm G}}) \sim \beta_0^{-2/3} v_{\rm rel}
(P_*/P_{\rm orb})^{3/8}$. Adopting the sound speed in a magnetized  ($\beta_0
\sim 1$) stellar wind in the neutron star vicinity to be  $\sim
10\,\text{km/s}$, we find that the angular velocity of the magnetic disk is
comparable to that of a neutron star in the present epoch provided the orbital
period of a binary system at the final stage of its evolution was in the range
$100 - 200$\,days.

 \section{Conclusions}\label{discussions}

The basic conclusion of our research is that the origin of an isolated X-ray
pulsar with the period of  6.7\,hr can be explained in terms of canonical
evolutionary scenario for a High-Mass X-ray Binary without an assumption about
the super-strong magnetic field on the neutron star surface. Our approach is
consistent with common belief about a relatively short initial spin period of
neutron stars and does not invoke additional assumption about a fall-back
accretion onto a neutron star after its birth. The age of 1E~1613 in our
scenario is in excess of a few million years, and the fossil disk, surrounding
the neutron star magnetosphere at the present time, was formed during its
evolution as a member of a HMXB.

According to our scenario of the origin of an isolated X-ray pulsar, the
neutron star turns out to be in the accretor state long before a supernova
explosion caused by the core-collapse of its massive companion. As a result,
the X-ray luminosity of such an object significantly exceeds its spin-down
power,
 $L_{\rm sd} = I \omega_{\rm s} \dot{\omega}$, that can be used as one of
 identification  criteria for this class of objects. The period of such a pulsar
 substantially depends on the parameters of a binary system, magnetic field strength on
 the neutron star surface and mass exchange rate between the system components.
However,  in all circumstances its value is limited as  $P_{\rm s} \geq P_{\rm
ma}$. Isolated pulsars with anomalously long periods are likely to be the
descendants of the widest pairs, in which the massive component underfills its
Roche lobe at the end of its evolution and the mass exchange between the
components occurs via a wind-fed accretion up to the moment of system
disruption. The life-time of such a pulsar is determined by both the mass of
residual disk and physical parameters in the gas defining the rate of magnetic
field diffusion. For the parameters of interest the dominant mode is an ohmic
diffusion with a characteristic time scale less than a few tens of thousands of
years. In this light, the best candidates for descendants of the high-mass
X-ray binaries are isolated X-ray pulsars associated with supernova remnants
which were formed in the process of core-collapse of their massive companions.
The objects of this type are known as  Compact Central Objects (CCOs), 1E~1613
is a representative of this class of isolated neutron stars.

\begin{theacknowledgments}
This work was partly supported by RFBR  grant No\,13-02-00077, SPbSU Grant
No\,6.38.669.2013 and the RAS Presidium Program No\,21 ``Non-stationary
phenomena in the Universe''.
\end{theacknowledgments}

\end{document}